\documentclass[12pt]{article}

\usepackage[utf8]{inputenc}
\usepackage[english]{babel}
\usepackage{amsmath}
\usepackage{amssymb}
\usepackage{graphicx,color}

\usepackage{multirow}

\usepackage{ulem}

\newcommand{\re}{Re}

\title{An example of the Rvachev function method}

\author{
{Alexander~V. Proskurin}\\
{\small \it Altai State University, Altai State Technical University, k210@list.ru}\\
Anatoly~M. Sagalakov\\
{\small \it Altai State University, amsagalakov@mail.ru}
}

\begin{document}

\maketitle

\begin{abstract}
We present a Rvachev function method with the Chebysev collocation for the stability analysis of fluid flow. The strategy is to construct an approximate solution that satisfies
all boundary conditions exactly. As an example, we consider the stability problem of the two-dimensional flow of an incompressible viscous liquid near a circular cylinder. The results coincide well with the reference data. The method is simpler than the widely used spectral/hp element method, in particular because it does not require mesh generation, and the collocation algorithm does not handle the boundary conditions or any geometric information.
\end{abstract}


\sloppy

\section{Introduction and problem formulation}


The development of new numerical methods that will be more efficient than existing ones is an important and interesting problem of computational fluid dynamics. A group of little-known methods for boundary value problems exists, based on Rvachev functions (see the review \cite{Shapiro:2007}). The Rvachev method allows picking out geometric and boundary conditions data from a numeric algorithm, which leads to significant simplifications.
Some examples of this method in fluid mechanics and a bibliography are presented in \cite{Shapiro:2003}.  

We consider the stability problem of an incompressible two-dimensional flow near a circular cylinder as a test case. This flow has been well-studied and the numerical results can be checked by comparison with the data from \cite{Barkley:2006,Theofilis:2009,Rogers:1988}. The problem of hydrodynamic stability is useful for the exploration of numerical methods, as it requires high quality approximation and yet is easier than time-dependent problems. Also, flow stability analysis plays an important role in understanding the process of a laminar flow's becoming turbulent.

The references \cite{Theofilis:2003,Theofilis:2011b} review the linear instability analysis of flows in complex two-dimensional (2D) and 3D geometries. This research area is called  global stability (or instability) analysis. These articles describe the methodology and main results of global stability analysis from the past three decades. It includes finite element and finite volume methods, finite differences methods, and spectral methods. High-order methods are strongly preferable when they allow of minimizing the computational cost. One of the best and most widely used techniques is the spectral/hp element method \cite{karniadakis:2013}, which employs meshes with rectangular or (and) triangular elements in complex domains. In each element of the computational domain, the solution is represented by a series of high order polynomials. Recent approaches in global stability analysis include: cavity and duct flows, 
flows near airfoils and cylinders, flows over steps and in corners, jet flows, etc.  

\begin{figure}[t]
\begin{center}
\includegraphics{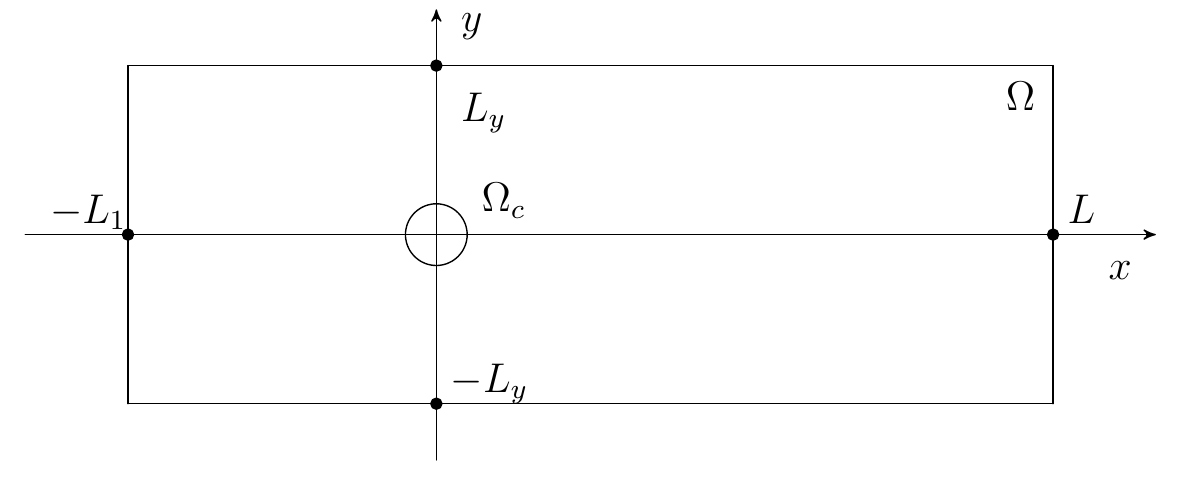}
\end{center}
\caption{Flow geometry setup.}
\label{flow_cyl}
\end{figure}

The method proposed below is simpler than the spectral/hp element method, in particular because it does not require generating  a mesh or the numerical algorithm's  handling either the  boundary conditions or any geometric information. Suppose the $x$-axis of a Cartesian coordinate system is collinear with the free-stream flow velocity and that the origin of this coordinate system matches is on the center line of a circular cylinder. The cylinder boundary is defined by $\Omega_c$ (see Fig. \ref{flow_cyl}). The $y$-axis is perpendicular to the $x$-axis. The computational domain is the rectangle $\Omega$ with dimensions $L$, $L_1$, and $L_y$.

The Navier--Stokes equation for the stream function $\Psi$ is
\begin{equation}
 \label{eq1}
\frac{\partial}{\partial t} \Delta \Psi - \frac{\partial \Psi}{\partial x}\frac{\partial \Delta \Psi}{\partial y}+\frac{\partial \Psi}{\partial y}\frac{\partial \Delta \Psi}{\partial x} = \frac{1}{\re}\Delta^2 \Psi,
\end{equation}
We define the Reynolds number as $\re=\dfrac{U_{\infty}d}{\nu}$, with $U_{\infty}$ being the magnitude of the free-stream flow velocity, $d$ the diameter of the cylinder, and $\nu$ the kinematic viscosity.

Our interest is in the evolution of infinitesimal disturbances of a base
flow. According to \cite{Henningson:2001,Theofilis:2003, Theofilis:2011b, Theofilis:2009}, the linearized Navier--Stokes equations governing these disturbances
are found by substituting
\begin{equation}
 \label{eq2}
\Psi=\Psi_0(x,y)+\psi(x,y)e^{Ct},
\end{equation}
into (\ref{eq1}) and keeping the linear terms. In (\ref{eq2}), $\Psi_0(x,y)$ denotes the base flow, $\psi(x,y)e^{Ct}$ is the infinitesimal disturbance, $\psi(x,y)$ the amplitude, $C=X+2\pi iY$, $X$  the growth rate, and $Y$ is the frequency of the disturbance. The resulting equations are
\begin{equation}
\label{eq3}
C \Delta \psi
=\frac{1}{\re}\Delta^2 \psi-V\Delta \psi_{y}+\psi_x \Delta U-U\Delta \psi_{x}+\psi_y \Delta V,
\end{equation}
where derivatives are denoted by subscripts, $U$ and $V$ are the $x$ and $y$ components of the base flow ($U=(\Psi_0)_y$, $V=-(\Psi_0)_x$), and $\Delta = \frac{\partial }{\partial x^2}+\frac{\partial }{\partial y^2}$.
The boundary conditions we consider are simple homogeneous on all boundaries as discussed in \cite{Barkley:2006,Theofilis:2009}: 
\begin{equation}
\label{eq4}
\psi = \frac{\partial \psi}{\partial x} = \frac{\partial \psi}{\partial y} = 0.
\end{equation}

\section{Methodology} 
A solution structure is a function that satisfies the boundary conditions exactly and contains the necessary degrees of freedom in order to approximate a solution of the problem. According to \cite{Shapiro:2003,Shapiro:2007}, for the considered problem, a solution structure can be written
\begin{equation}
 \label{eq5}
\psi=\omega^2 \sum_{i=0}^n \sum_{j=0}^k a_{ij}T_i(x)T_j(y),
\end{equation}
where $\omega(x,y)$ is the boundary distance function, $T(x)$ and $T(y)$ are Chebyshev polynomials of the first kind on appropriate intervals, and the $a_{ij}$ are unknown coefficients. The set of collocation points is chosen as the set of zeros of the Chebyshev polynomials. We posit that (\ref{eq5}) satisfies the equation (\ref{eq3}) at the set of collocation points and get the algebraic eigenvalue problem
\begin{equation}
\label{eq6}
 \boldsymbol{A}\boldsymbol{v} = C \boldsymbol{B} \boldsymbol{v},
\end{equation}
where $\boldsymbol{v}=\{ a_{00}, a_{01},\ldots, a_{n(k-1)}, a_{nk} \}$.

 So we carry out the discretization in coefficient space.

Iteration methods implemented by the {\bf SLEPc} software package \cite{slepc_manual} have been used for the eigenvalue problem (\ref{eq6}). We compute the part of spectrum with the largest growth rates.

The boundary distance function $\omega_{\Gamma}$ is considered as a function which should only be equal to zero on the boundary $\Gamma$. This function changes sign only on the boundary. Obviously, the boundary function may be guessed for simple geometrical figures such as lines or circles. For complex domains, this function is constructed by algebraic combination (conjunction or disjunction) of more simple figures by $R$-operations (see works\cite{Shapiro:2003,Shapiro:2007,Rvachev:1982}). For example, $R$-conjunction makes

\begin{figure}
\begin{center}
\includegraphics{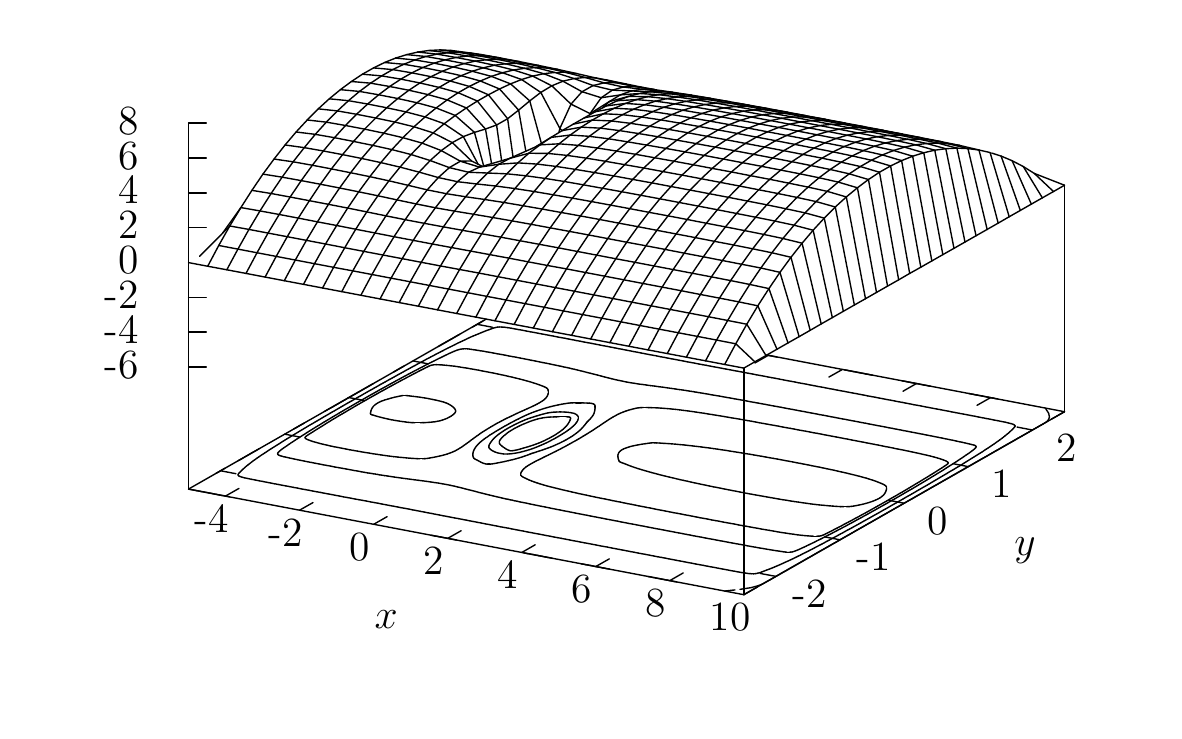}
\end{center}
\caption{The boundary distance function.}
\label{b_dist_func}
\end{figure}

\begin{equation}
\label{eq7}
\omega_{\Omega}(x,y) = x_{S1} \wedge_{\circ} x_{S2} \equiv x_{S1}+x_{S2}-\sqrt{x_{S1}^2+x_{S2}^2},
\end{equation}
where the boundary distance function of rectangle $\omega_{\Omega}(x,y)$ has been obtained from the functions of two stripes $x_{S1}=k_1(L_y^2-y^2)$ and $x_{S2}=k_2(x-L)(L_1-x)$. For the cylinder, we consider the boundary distance function as $\omega_{\Omega_C} = -k_3(x-L-10)(x^2+y^2-0.25)$ and use the $R$-conjunction (\ref{eq7}) again(see Figure \ref{b_dist_func}):
\begin{equation}
\begin{aligned}
\label{eq7.1}
k&=\sqrt{1.0 \times 10^{-4}\,\left(6.25-y^2\right)^2+0.01\,\left(x-10\right)^2\,\left(x+5\right)^2}\\
\omega&=k_4\,\left(-\left(\left(-k+0.01\,\left(6.25-y^2\right)-0.1\,\left(x-10\right)\,\left(x+5\right) \right)^2+ \right. \right. \\
 &+ \left. 1.0 \times 10^{-4}\,\left(20-x\right)^2\,\left(y^2+x^2-0.25\right)^2\right)^{\frac{1}{2}}-k+0.01\,\left(20-x \right)\times  \\
 &\times \left. \left(y^2+x^2-0.25\right)+0.01\,\left(6.25-y^2\right)-0.1\, \left(x-10\right)\,\left(x+5\right)\right).
\end{aligned}
\end{equation}
Here, $k_1=k_3=0.01$, $k_2=0.1$, and the common multiplier $k_4=100$.

For analytical computations of boundary functions we use the {\bf Maxima} computer algebra system. This method allows us to manipulate simple geometry, but it is adequate for our method demonstration with only standard software. The methods of calculation of the boundary distance function are reviewed in work \cite{Shapiro:2002}, as it states that such a function may be calculated fully automatically for arbitrary complex geometry using the special software { \bf SAGE} \cite{Shapiro:2002} and {\bf POLE} \cite{Rvachev:1982}.

The base flow was computed using the spectral/hp element framework {\bf Nektar++}, but any flow solver may be used instead, for example, {\bf OpenFoam}, {\bf CodeSaturne}, etc.   The dimensions of the domain are taken to be $L=100$, $L_1=20$ and $L_y=10$. A uniform flow $(U = U_{\infty},\,V=0)$ is imposed at the inflow, top, and bottom boundaries; no-slip conditions are imposed on the cylinder surface. A zero traction condition is imposed at the outflow boundary of the domain. The base flow can be stabilized by setting $V=0$ at $y=0$ in the cylinder trace as recommended in \cite{Barkley:2006,Theofilis:2009}. An unstructured triangle mesh was obtained from the grid generator {\bf gmsh}. The flow fields were verified by data from the review \cite{Rogers:1988}.

\section{Results and discussion}

We have proved the method for the example of the stability problem of plane Pouiseuille flow \cite{Proskurin:2013}. In this work, the authors have compared his method  with  the spectral/hp element method using the {\bf Nektar++} framework. These two approaches are similar in terms of precision and calculation time, but not in terms of memory volume. Our method requires a high volume of memory for the big dense matrices $A$ and $B$.

The size of the domain is taken to be smaller than that for the base flow: $L=50$, $L_y=7.5$, and $L_1=10$. We compute growth rates $X$ for some range of $n$ and $k$ in order to verify and validate the method. The results are presented in Table \ref{tab11} for $\re=45$, $\re=46$, $\re=47$, $\re=50$.

\begin{table}[h]
\caption{Dependence of growth rate $X$ on $n$ and $k$ at $\re=45$, $\re=46$, $\re=47$, $\re=50$}
\label{tab11}
\begin{center}
\begin{tabular}{|c|c|c|c|c|c|c|}  
\hline
$n$&40&80&100&180&\multirow{2}{*}{Nektar++}&\multirow{2}{*}{$\re$}\\
\cline{1-5}
$k$&20&40&50&60& &\\

\hline

\multirow{4}{*}{$X$}&0.4088&0.0872&0.0010&\uline{-0.0037}&-0.0024&45\\
&0.4022&0.0868&\uline{0.0020}&0.0150&0.0020&46\\
&0.3408&0.0852&\uline{0.0066}&0.0040&0.0066&47\\
&0.3958&0.0490&0.0106&\uline{0.0170}&0.0170&50\\
\hline
\end{tabular}
\end{center}
\end{table}

The conclusion based on the data in Table \ref{tab11} is that increasing the number of modes above some limit does not lead to an increase in precision. We assume that the cause of the lost precision is errors at the stage of the  algebraic eigenvalue problem. Methods for large dense algebraic eigenvalue problems are still not sufficiently investigated and are frequently unstable.

\begin{figure}
\begin{center}
\includegraphics{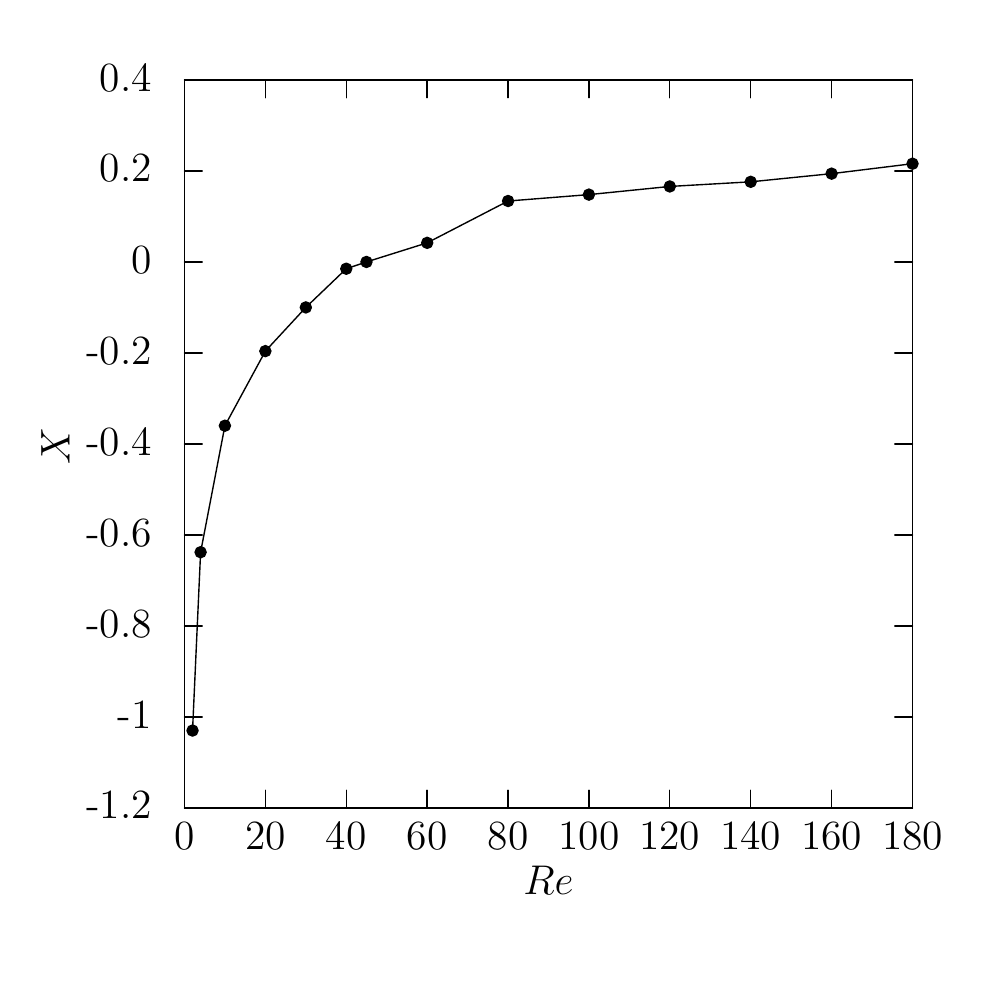}
\end{center}
\caption{Growth rates as a function of Reynolds number}
\label{X_Re}
\end{figure}

\begin{figure}
\begin{center}
\includegraphics{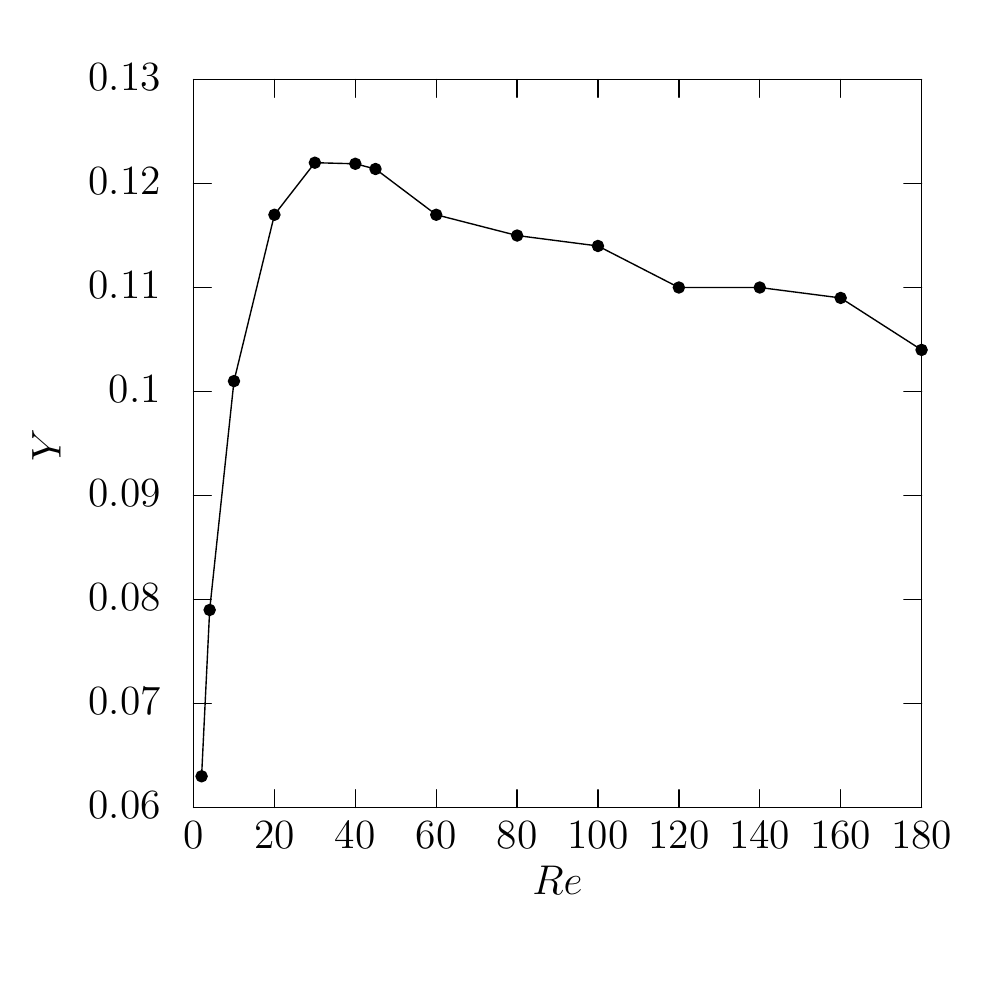}
\end{center}
\caption{Frequencies as a function of Reynolds number}
\label{Y_Re}
\end{figure}

\begin{figure}
\begin{center}
\includegraphics{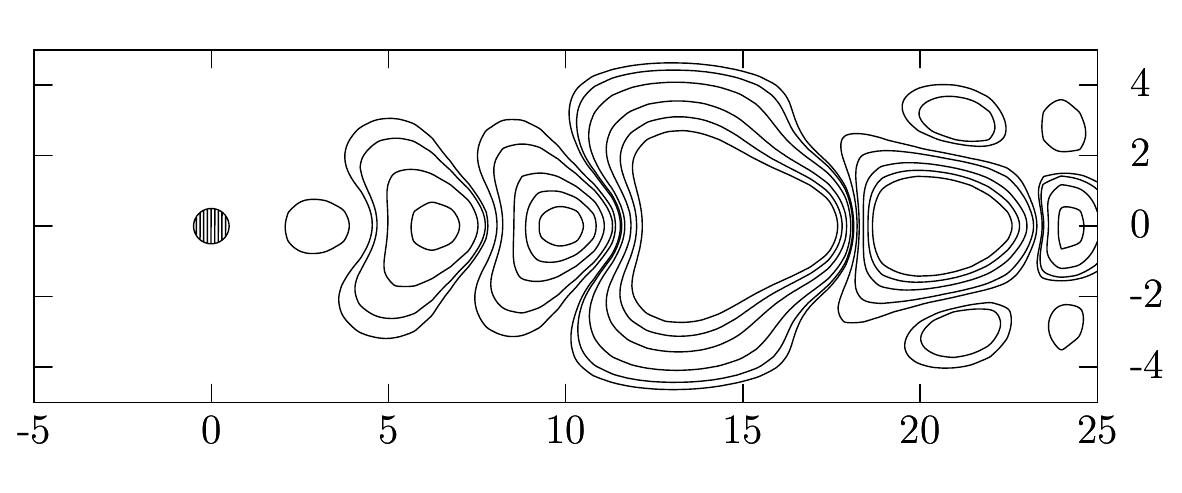}
\end{center}
\caption{Streamlines of the real part of leading eigenmode at $\re=30$}
\label{ef_Re15}
\end{figure}

\begin{figure}
\begin{center}
\includegraphics{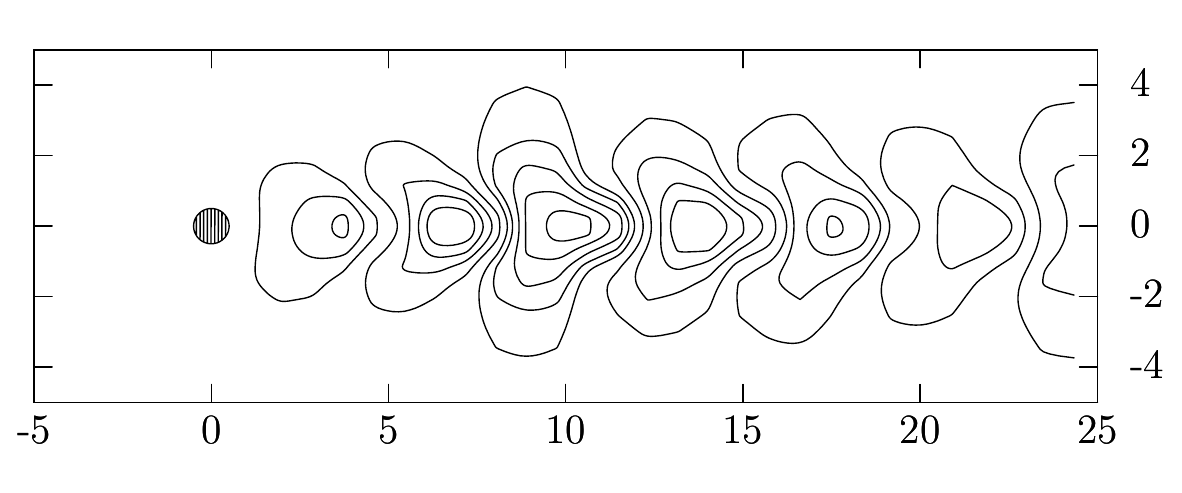}
\end{center}
\caption{Streamlines of the real part of leading eigenmode at $\re=80$}
\label{ef_Re40}
\end{figure}

Figures \ref{X_Re} and \ref{Y_Re} show the growth rates and frequencies as functions of the Reynolds number($L=50$, $L_y=5$, and $L_1=10$; $n=200$ and $k=60$). This result matches the data from \cite{Barkley:2006}. The number of collocation points is on the $x$-axis is $n=200$, and on the $y$-axis, $k=60$.  The critical Reynolds number lies between 45 and 46, which is close to the result from \cite{Barkley:2006, Theofilis:2009}. Figures \ref{ef_Re15} and \ref{ef_Re40} show the streamlines of the real part of the leading eigenmode at $\re=30$ (stable) and $\re=80$ (unstable), $L=25$, $L_1=5$ and $L_y=5$.

So we can conclude that the new method has an error value close to that of the spectral/hp-element method, but is simpler and allows problems similar to those  in the reviews \cite{Theofilis:2003,Theofilis:2011b} to be solved. The simplicity of the method lies in the separation of geometric data from the computation algorithm. The solution structure (\ref{eq5}) was determined once before the start of calculations and contains the boundary conditions and shape of the domain. Further ones can construct the algorithm as the usual spectral scheme. The method does not directly manipulate the geometric data and boundary conditions at this stage and can be realized more simply than the finite difference and finite element methods. Of cource, when including $R$-function theory and solution structure construction algorithms,  the method  is more complicated.

\bibliographystyle{plain}
\bibliography{reference1}

\newpage

\end{document}